\documentclass{article}

\usepackage{ijcai09}
\usepackage[safe]{tipa}
\usepackage{calc}
\usepackage{amstext}
\usepackage{times}
\usepackage{helvet}
\usepackage{courier}
\usepackage{amsmath, amssymb}
\usepackage{latexsym}
\usepackage{eurosym}
\usepackage{wasysym}
\usepackage{cancel}

\makeatletter
\newif\if@restonecol
\makeatother

\usepackage[ruled,lined,boxed]{algorithm2e}
\usepackage{tikz}
\usepackage{subfigure}
\usepackage{multicol}
\usepackage{pslatex}
\usepackage[small]{caption}
\usepackage{multirow}
\usepackage[arrow, matrix, curve]{xy}
\usepackage{epsfig}
\usepackage[pdfmark]{thumbpdf}
\begin{document}
\graphicspath{{../all/eps/}}

\title{General Game Management Agent}
\author{Paper ID: 2\\
Institute for my obsession}
\maketitle
\begin{abstract}
The task of managing general game playing in a multi-agent system is the problem addressed in this paper. It is considered to be done by an agent. There are many reasons for constructing such an agent, called general game management agent. This agent manages strategic interactions between other agents - players, natural or also artificial. The agent records the interaction for further benchmarking and analysis. He can also be used for a kind of restricted communications. His behavior is defined by a game description written in a logic-based language. The language, we present for this application, is more expressive than the language GDL, which is already used for such purposes. Our language can represent imperfect information and time dependent elements of a game. Time dependent elements like delays and timeouts are of crucial importance for interactions between players with bounded processing power like humans. We provide examples to show the feasibility of our approach. A way for game theoretical solving of an interaction description in our language is considered as future work.
\end{abstract}
\section{Introduction} 
\indent If rational agents interact, they pursuit their interests. It is not surprising, because rational agents persuit their interests all the time. If agents are advanced enough, they know or try to find out the interests of other agents and that their interests are also known or tried to be found out and so on. It is called strategic interaction. The common knowledge of rationality between the agents is one of the base assumptions needed for classical game theoretic analysis. In game theory, we have the notion of equilibrium. An equilibrium is a combination of behaviors, none of whose owners is interested in deviating from. If a game (used as synonym to strategic interaction) has a finite number of participants, their action and states, it is finite. Infinite games are difficult to model for computation. This work considers only finite games. A finite game has at least one (mixed strategies) equilibrium \cite{nash}. Finding the equilibria is solving games. But, the calculating of the exact solution costs time. In most real-life domains, agents have not enough time or they are not advanced enough to do it properly. Real agents are supposed to behave suboptimally. For instance, humans deviate even in very primitive games like Roshambo significantly from equilibria \cite{framasi01}.\\
\indent In Artificial Intelligence, games are interesting from at least two points of view - mechanism design and agent design \cite[p.632]{russel}. In mechanism design, we want to achieve a kind of (cooperative) behavior and need rules. For agent design, we have rules and try to derive optimal behavior. In most cases, the system represents one special game like chess and has to compute the best strategy as fast as possible. In such cases, the game is represented by a pile of low level code. In practical view, every agent has to represent the game and then a component is required which represents the game for all agents. In global view, this required component is a (game) server, which implements rules for interaction. This means, that one has to solve in general two tasks for game computing - game server and game solver. Game solver is the part used by agents. There are many techniques for computational solving games - from optimal game theoretic to suboptimal AI heuristics. If we have a library with code for chess game server and a library for an agent, which does his best in playing chess, it does not mean that we can use this libraries for other games. Additionally, there is a hard manageable game representing code on both sides. The low-level code based game server can not send the game rules to the players, unless he sends his code. The game rules are typically reflected in a network communication protocol for the server. Consequently, if an agent does not have a representation of the game, he can not participate properly in it. And we can not describe in a clean way a behavior of an agent as depending on rules of the game, because these rules are hard-coded.\\
\indent In next section we provide basic concepts for this paper. Then, we present the definition of our strategic interaction definition language (SIDL\footnote{\textipa{["zaId@l]}}). In section \ref{beyond}, we describe the algorithms around SIDL. Section \ref{case} present some application examples. After it, we conclude and look forward.
\section{Preliminaries}
\indent To get rid of the problems described in the previous section, one has to develop a high level language to define games. Summarized, it causes following advantages:
\begin{itemize}
\item A clear and convertible representation of a game is given. Convertible means that it can be easily converted to other representation formats and formalisms. Such representations can be exchanged between systems - from a game server to a player e.g..
\item The game solver and the game server use the same file. This reduces redundancy. It also guarantees that both run on the same game.
\item The rules can be edited easier. One can also develop algorithms, which manipulate game representation in any possible way - 'remove simultaneous turns' or 'reduce number of players' e.g..
\item One can develop techniques for defining behaviors of players on the basis of a game description. Players which are defined for playing multiple kinds of games can be constructed and benchmarked.
\end{itemize}
The only disadvantage can be the computation time. Low level code solutions for specific games can be done slighter \cite{ggp}. That saves computation time.\\
\indent The idea of a game description language is already partially considered in GALA-System \cite{gala} and GGP \cite{ggp}. The GALA-System is a high level language based game solver. Solving of games in GALA is based on a state-of-art game theoretic software GAMBIT \cite{gambit}. The GALA-Language is a logic based language, which enables a prolog interperter to generate huge game trees for GAMBIT. Game trees for GAMBIT are huge, because they have inter alia repeated states. GGP is about a high level language based game server using the logic based language GDL. It is used for AI programming contests. GALA and GGP are independent. The language PNSI \cite{tagiewcimca} is able satisfy both tasks - game server and solver. PNSI is based on Petri Nets. Unfortunately, Petri Nets based representation is not so slight as a logic based approach. Using logic, we can define game rules more generally.\\
\indent We consider the general game management as a task of an agent and not as a server. Our definition of the game management agent\footnote{unequal to federal game management agent} (GMA) is similar to the definition of the world and market agents in the computational economics framework ACE \cite{tesf}. The world agent manages the state of the environment and the market agent manages prices. General GMA (GGMA) is a GMA, which is based on general game description language. The goal of our GGMA is in general 'producing' agents's strategic interaction. To 'produce' means here to let it happen. For other agents, it means that GGMA is interested in providing strategic interaction with other agents. GGMA records and can analyze the behavior of the players. In special cases, one can add a couple of constraints about the behavior, which is required to be produced. Further, one can use GGMA for agent and mechanism coevolution \cite{coevolution}. GGMA is the way for definition of mechanism design as task of an agent. It is much easier to define mutation in game description, which is written in logic as in C++. Otherwise, GGMA can be used as an usual game server for conducting experiments \cite[e.g.]{framasi01}.\\
\indent Let us go back now to the question of the desired expressiveness of SIDL. It must represent at least finite games. For instance, GALA can represent finite games of imperfect information and GDL only of perfect information. Games of incomplete\footnote{unequal to imperfect} information can be transformed to games of imperfect information \cite{rubin}. GAMBIT is able to solve games of imperfect information. One another aspect in games, which is not considered in classical game theory, is time. That is why the prefix of our language is strategic interaction rather than game. Time in games is very important in real-life domains. Delays and timeouts can extend or reduce computation time for decision making and hence change the resulting decision. It is also interesting for the behavior analysis, at which time a decision is made. In many cases, agents reason about the time of other agents needed to make a decision. Timing is often of crucial importance. SIDL is considered to represent finite games of imperfect information with discrete time.\\
\section{SIDL Definition}
\indent SIDL is strongly related to GDL. GDL provides as game model, which is a game graph. This game graph has no repeated states unlike the game trees for games in extensive form. Every node of this graph is a game state described by a couple of facts. A game state in GDL is not monolithic. It can be manipulated by logic operations like a database. GDL uses following key words:\\
\begin{description}
\item[$role(R)$]: A participating agent $R$.
\item[$init(P)$]: A fact $P$, which is true at the initial state.
\item[$true(P)$]: A fact $P$, which is true at the current state.
\item[$next(P)$]: A fact $P$, which is true at the next state.
\item[$legal(R, M)$]: An action $M$, which is legal for a participant $R$.
\item[$does(R, M)$]: A player $R$, does an action $M$.
\item[$goal(R, V)$]: A player $R$ gets an value $V$.
\item[$terminal$]: terminal holds, if the current state is terminal.\\
\end{description}
\indent The rules for manipulation of the current state or also a pile of facts are written in logic. For the initial state $true(P)$ holds, if $init(P)$ holds. The difference between $init$ and $true$ is that $true$ appears only as precondition for a rule. In contrast, $init$ is a static statement. The postcondition of logic rules are facts held by the predicates $next$, $legal$, $goal$ and $terminal$. The facts held by $does$ are the commands sent from the players to the game server. For every fact, which must hold in the next state, one must define a rule headed by $next$. Rules with the predicate $legal$ prevent players from creating facts held by predicate $does$, which are not in terms of the game definition. GDL can represent only finite games of perfect information, because there is no mechanism to hide information from agents's eyes or to enable simultaneous moves. There are also no time dependent elements. \\
\indent Regarding the deficiencies of GDL, we decided to develop a particular philosophy for such a language. A strategic interaction is a kind of a running engine. The engine runs in discrete 'steps'. The time period between two 'steps' is always the same and called chronon. After a chronon is expired, the engine makes some changes in its internal state. This engine has a couple of switches. The set of switches is constant. Every switch has a current state and an owner, who can alter its state. Switching of the switches can impact the running of the engine. A player acts by switching a switch. Some player are able to see some of details of the internal state of the engine.\\
\indent Based on this philosophy, we constructed our language SIDL:\\
\begin{description}
\item[$chance(<BID>, <Distribution>)$]: A distribution for random turns. $<BID>$ is branching ID. $<Distribution>$ is an array of real numbers. Sum of $Distribution>$ is $1.0$. Every member of $<Distribution>$ is higher than $0.0$ and smaller than $1.0$.  
\item[$switch(<BID>, <Agent>, <Aliases>)$]: A switch owned by a player $<Agent>$ with an array of actions $<Aliases>$. $<BID>$ is unique. A $<BID>$ identifies either a chance or a switch.
\item[$fact(<P>, <Arity>)$]: Definition of a fact, which can be stored in the database. $<P>$ is a predicate name, which is not in the set of key words for SIDL.
\item[$hidden(<F>, <HiddenFor>)\Leftarrow(\bigwedge \limits_{0}^{n} <F>)$]: A rule for definition of imperfect information, where $<F>::=<P>(...)$. $<HiddenFor>$ is an array of agents, for which the fact is hidden. Using preconditions for this rule, one can exactly define, what every agent can see and what not.
\item[$init(<F>)$]: This has the same meaning as in GDL. 
\item[$init(account(<Agent>, <Real>))$]: $account$ is a key word in SIDL. It denotes an account balance of an agent.
\item[$init(does(<BID>, <Alias>))$]: $does$ is a also a key word in SIDL. It denotes a state of a switch.
\item[$operation(<Operator>)\Leftarrow$]
$((\bigwedge\limits_{0}^{n} <F>) \wedge (\bigwedge\limits_{0}^{m} <A>)\wedge$\\
$(\bigwedge\limits_{0}^{l} <N>)\wedge (\bigwedge\limits_{0}^{k} <G>))$
:$\;\;$ Definition of a state manipulation operation. It can have four kinds of preconditions. It is supposed to be lazy evaluated. $<F>$ is for testing the state. If $<F>$-statements hold, the rule manipulates the state using the side effect predicates denoted below.
\item[$<A>::=ax(<F>)$]: Removes immediately a fact.
\item[$<N>::=next(<F>)$]: Creates a fact in the next state
\item[$<G>::=goal(<Agent>, <Payoff>)$]: Adds $<Payoff>$ to the account balance of $<Agent>$. 
\item[$branching(<Operators>, <BID>)$]: $<Operators>$  is an array of operations, from which only one can be executed in a chronon. A branching is related to a switch or a chance. The length of the aliases or the distribution in the related switch or chance must be exactly the same as the length of operators in branching. Multiple branchings can be related to the same switch or chance. Branchings with only one operation have value $nil$ as $BID$.
\item[$command(<Agent>, <BID>, <Alias>)$]: This is performed, if a player sends his action to the GGMA. The side effect is that the related $does$-fact is reassigned.
\item[$terminal\Leftarrow(\bigwedge \limits_{0}^{n} <F>)$]: This has the same meaning as in GDL. 
\end{description}
\indent The facts are dynamic as in GDL. SIDL has no rules headed by $next$. $next$ is used for precondition statements and has a side effect. The side effect is that the fact held by $next$ is holding in the next state. A fact is kept in the database till it is not removed using predicate $ax$. We abandon the predicate 
$role$. $role(X)$ is replaced by $init(account(X,\_))$. We do not need the predicate $legal$. That is because of our philosophy. It is never illegal to switch own switches, but it does not always cause any effect. The game terminates, if the rule $terminal$ holds. Multiple branchings can be related to one switch. That is for modelling imperfect information. In case of imperfect information, an action has different consequences depending on current state which known to the player.\\
\section{Beyond Definition}
\label{beyond}
\indent The question, we handle in this chapter, is how to run a GGMA on a game description in SIDL. GGMA promotes his game management service for a game described in SIDL. The other agents analyze the game definition and decide to participate or not. After GGMA gathers enough players, he starts to execute Alg. \ref{routine} which a combination of imperative code and logic definitions. The names of the methods in this algorithm are self-descriptive. The function 'lrun' is a connection between the imperative code of GGMA and a logic interpreter for SIDL. It returns all possible assignments of variables of a logic statement. The logic operation $chronon$ runs over all branchings, chooses an operation and executes it if possible. A definition for $chronon$ is below. The rule $tryoperation$ expresses the fact that in a game state only a subset of operations can be executed. The logic operation $request$ deliveres all facts and for every fact names of agents barred  from viewing it:\\
\begin{flushleft}
$tryoperation(O)\Leftarrow operation(O)\vee true$\\
$handleBr(Operators, BID)\Leftarrow$\\
$\;\;(chance(BID, D) \wedge random(Operators, D, O) \wedge$\\
$\;\;\;\;tryoperation(O)) \vee$\\
$\;\;(switch(BID, \_, As) \wedge$\\
$\;\;\;\; does(BID, A) \wedge, map(As, A, Operators, O) \wedge$\\
$\;\;\;\; tryoperation(O))$\\
$chronon \Leftarrow$\\
$\;\; \forall Operators (branching(Operators, BID) \rightarrow$\\
$\;\;\;\; handleBr(Operators, BID))$\\
$request(X, H) \Leftarrow X \wedge hidden(X, H)$\\
\end{flushleft}
\indent 
\begin{algorithm}
\incmargin{1em}
\linesnumbered
\caption{GGMA execution}
\label{routine}
\SetLine
\dontprintsemicolon
\KwData{SIDL}
\While {not lrun(terminal)} {
\While{not a\_chronon\_expired}{
command = receive\_command\_for\_altering\;
lrun(command(command.agent,command.bid, command.alias))\;
record(command)\;
}
lrun(chronon)\;
state = lrun($(X, H) : \forall X request(X, H)$)\;
accounts = lrun($(A, M) : \forall X account(A, M)$)\;
record(state, accounts)\;
send2agents\_regarding\_hidden(state, accounts)\;
}
\end{algorithm}
\section{Case Study}
\label{case}
\indent To demonstrate SIDL in practice, we provide some examples. As we claimed in contrast to GDL, SIDL has to be able to represent two additional features - imperfect information and time dependent elements. For the first feature, we considered following situation. After a turn of the nature, Alice does not know the current state, in which she has two actions 'A', 'B' and 'Wait'. There are only two states possible. This is a very simple situation with imperfect information. We represent it in following prolog style code:
\footnotesize
\begin{verbatim}
fact(state, 1). // state('current state ID')
request(state(X), [alice]):-
   state(X).
branching([nat(1), nat(2)], 0).
branching([a(1), b(1), wait], 1).
branching([a(2), b(2), wait], 1).
chance(0, [0.5, 0.5]).
switch(1, alice, ['A', 'B', 'Wait']).
operation(nat(X)) :- 
   state(0),
   ax(state(0)),
   next(state(X)).
operation(a(X)) :- 
   state(X),
   ax(state(X)),
   next(state(10)),
   goal(alice, 3-X). // payoff function
operation(b(X)) :- 
   state(X),
   ax(state(X)),
   next(state(10)),
   goal(alice, X). // payoff function
operation(wait) :-
   false.
terminal :-
   state(10).
init(account(alice, 0.0)).
init(does(1, 'Wait')). // waiting at start
init(state(0)).
\end{verbatim}
\normalsize
\indent As you see, Alice can not see the current state of the game. The operations of only one of the branching with $BID=1$ are executable at same time. Alice can not change the executing operation for every of her branchings separately. Operation $wait$ causes nothing.\\
\indent A time period in a strategic interaction using SIDL is an amount of chronons. If the chronon is defined to be $1$ sec, one can only act with time period of $1$ sec, $2$ sec, $3$ sec and so on. If we want to define a delay of $30$ sec, we initialize a fact like $countdown(30)$. Then we define a branching with only one rule. The rule in this branching decrements $X$ in $countdown(X)$. In a same way, we can define a timeout. Sudden events can be modelled by a chance, which is related to a branching with a $wait$ and a $suddenEvent$ operation.\\
\indent The last example is a game with simultaneous turns. We take for this example a single turn in the game 'Pico 2' \cite{tagiewmnm}. There are 11 cards - $4$ till $13$ and $16$. A card beats another card, if it is higher and not higher than two times the lower card. Following SIDL code is a reduced representation of a single turn in this game. 
\footnotesize
\begin{verbatim}
fact(onhand, 2). // onhand('Agent','a card of his')
fact(thrown, 2). // thrown('Agent','his thrown card')
branching([a(4),a(5),a(6),a(7),a(8),a(9),
   a(10),a(11),a(12),a(13),a(16)], 1).
branching([b(4),b(5),b(6),b(7),b(8),b(9),
   b(10),b(11),b(12),b(13),b(16)], 2).
branching([payoff], nil). // BID is not required
switch(1, alice, ['4','5','6','7','8','9',
   '10','11','12','13','16']).
switch(2, bob, ['4','5','6','7','8','9','10',
   '11','12','13','16']).
operation(a(X)) :- 
   onhand(alice, X), // alice has a card X
   not(thrown(alice, \_)), // alice did nothing yet
   ax(onhand(alice, X)), // remove X from hand
   next(thrown(alice, X)). // create thrown X
operation(b(X)) :- 
   onhand(bob, X),
   not(thrown(bob, \_)),
   ax(onhand(bob, X)),
   next(thrown(bob, X)).
operation(payoff) :- 
   thrown(Agent1, C1), // thrown cards
   thrown(Agent2, C2),
   C1 > C2, // game rule definition
   C1 <= C2*2,
   goal(Agent1, 1). // Agent1 won
\end{verbatim}
\normalsize 
\section{Conclusion}
\indent In this work, we discussed the advantages of using of a description language for general strategic interactions. The related work about this theme is summarized. A language SIDL is presented. SIDL transcends in its expressiveness all previous approaches. Examples for SIDL files are given. Further, a design of GGMA is presented. GGMA is an innovative approach in game computing. GGMA is an agent based manifestation of mechanism design intent.\\
\indent As future work we consider an automated method for constructing GAMBIT acceptable game trees on the basis of a SIDL file. This makes possible to solve analytically game descriptions made in SIDL. The other direction is visualization of a SIDL file and also visual editing.\\
\bibliographystyle{named}
\bibliography{sidlgiga}

\begin{thebibliography}{}

\bibitem[\protect\citeauthoryear{Genesereth \bgroup \em et al.\egroup
  }{2005}]{ggp}
Michael~R. Genesereth, Nathaniel Love, and Barney Pell.
\newblock General game playing: Overview of the aaai competition.
\newblock {\em AI Magazine}, 26(2):62--72, 2005.

\bibitem[\protect\citeauthoryear{Koller and Pfeffer}{1997}]{gala}
Daphne Koller and Avi Pfeffer.
\newblock Representations and solutions for game-theoretic problems.
\newblock {\em Artificial Intelligence}, 94(1-2):167--215, 1997.

\bibitem[\protect\citeauthoryear{Nash}{1951}]{nash}
John Nash.
\newblock Non-cooperative games.
\newblock {\em Annals of Mathematics}, (54):286 -- 295, 1951.

\bibitem[\protect\citeauthoryear{Osborne and Rubinstein}{1994}]{rubin}
Martin~J. Osborne and Ariel Rubinstein.
\newblock {\em A course in game theory}.
\newblock MIT Press, 1994.

\bibitem[\protect\citeauthoryear{Phelps}{2007}]{coevolution}
S.~G. Phelps.
\newblock {\em Evolutionary Mechanism Design}.
\newblock PhD thesis, PhD thesis, University of Liverpool, 2007.

\bibitem[\protect\citeauthoryear{Russel and Norvig}{2003}]{russel}
Stuart Russel and Peter Norvig.
\newblock {\em Artificial Intelligence}.
\newblock Pearson Education, 2003.

\bibitem[\protect\citeauthoryear{Tagiew}{2008a}]{tagiewcimca}
Rustam Tagiew.
\newblock Multi-agent petri-games.
\newblock In Masoud Mohammadian, editor, {\em CIMCA08 -- IAWTIC08 -- ISE08 (in
  press)}. IEEE Computer Society Publishing, 2008.

\bibitem[\protect\citeauthoryear{Tagiew}{2008b}]{tagiewmnm}
Rustam Tagiew.
\newblock Simplest scenario for mutual nested modeling in
  human-machine-interaction.
\newblock In {\em Advances in Artificial Intelligence}. Heidelberg, Springer,
  2008.

\bibitem[\protect\citeauthoryear{Tagiew}{2009}]{framasi01}
Rustam Tagiew.
\newblock Towards a framework for management of strategic interaction.
\newblock In {\em First International Conference on Agents and Artificial
  Intelligence}, pages 587 -- 590. INSTICC, 2009.

\bibitem[\protect\citeauthoryear{Tesfatsion and Judd}{2006}]{tesf}
Leigh Tesfatsion and Kenneth Judd.
\newblock {\em Handbook of computational economics}, volume~2.
\newblock Elsevier/North-Holland, 2006.

\bibitem[\protect\citeauthoryear{Turocy}{2008}]{gambit}
Theodore~L. Turocy.
\newblock Toward a black-box solver for finite games.
\newblock In {\em IMA Software for Algebraic Geometry Workshop}. Springer,
  2008.

\end{thebibliography}
\end{document}